\magnification 1200
%%
%% STANDARD MACRO STUFF
%%

\font\titlefont=cmss17 at 29.88 true pt
\font\authorfont=cmssi17 at 20.74 true pt
\font\bigaddressfont=cmss12 at 14.4 true pt 
 at 14.4 true pt
\def \mysubmit {}
\def \mypresent {}

\def \docnum #1 { \def \mydocnum {#1}} 
\def \date #1 { \def \mydate {#1}} 
\def \title #1 {\def \mytitle {#1}}
\def \author #1 {\def \myauthor {#1}}
\def \abstract #1 {\def \myabstract {#1}}

\def \tobesubmittedto #1 { \def \mysubmit {\leftskip=0pt plus 1fill \rightskip=0pt plus 1fill 
\hbox{\vbox{\noindent \hfill \it To be 
submitted To #1 \hfill \ }}}}
\def \submittedto #1 { \def \mysubmit {\leftskip=0pt plus 1fill \rightskip=0pt plus 1fill 
\hbox{\vbox{\noindent \hfill \it Submitted 
To #1 \hfill \ }}}}
\def \presentedat #1 { \def \mypresent {\leftskip=0pt plus 1fill \rightskip=0pt plus 1fill 
%\hbox{\vbox{\noindent \hfill \it Presented 
\hbox{\vbox{\noindent \it Presented 
at #1  }}}}

\def\maketitle{
\let\footnotesize\small
\let\footnoterule\relax

\ifx\mydate\undefined \def \mydate {
\ifcase\month\or
January\or February\or March\or April\or May\or June\or
July\or August\or September\or October\or November\or December\fi
\space\number\day, \number\year} \fi

\ifx\thispagestyle\undefined \nopagenumbers \fi
\ifx\nopagenumbers\undefined {\thispagestyle{empty}} 
      \setcounter{page}{0}%
      \fi
\null
\rightline{\logo}
\vskip 2 pt
\rightline{\mydocnum}
\rightline{\mydate}
\vskip 20 pt
{\def\\{\break} 
\leftskip=0pt plus 1fill
\rightskip = 0 pt plus 1fill
\parindent 0 pt
\baselineskip 30 pt
\titlefont \hfil \vbox { \mytitle}\hfil }
\vskip 25 pt
{\def \and {\qquad} 
\leftskip=0pt plus 1fill
\rightskip = 0 pt plus 1fill
\parindent 0 pt 
\authorfont
\lineskip 12 pt
\myauthor
\parfillskip=0pt\par
}%
\vskip 20 pt

{\bigaddressfont
\centerline{ Department of Physics}
\centerline{ Manchester University}
\centerline{ England}
}

\ifx\myabstract\undefined {}
\else
\null\vfil\vskip 10 pt
\centerline{ \bf Abstract}
\vskip 10 pt
\myabstract
\fi

\ifx\footline\undefined   % LATEX stuff
\begin{figure}[b]
\mypresent
\mysubmit
\end{figure}
\mythanks
\setcounter{footnote}{0} 
\vfil
\null
\else                       %Plain TeX stuff
\footline={{\baselineskip=10 pt \vbox{\hbox to \hsize {\mypresent} \vskip 5 pt \hbox to \hsize{\mysubmit}}}}
\vfil
\eject
\pageno=1 % title page is not given a page number
\footline={\hss\tenrm\folio\hss}
\fi
				      
\let\thanks\relax
\gdef\mysubmit{}\gdef\present{}
\gdef\mythanks{}\gdef\myauthor{}\gdef\@title{}\let\maketitle\relax}

\def \phonenumber{4170}
\def\today{\number\day/\number\month/\number\year\space\number\hour%
:\number\minute\space\jobname}

\def \logo {\vbox to 23.5 mm {
\hbox {}
\hbox to 47 mm 
{
\includegraphics{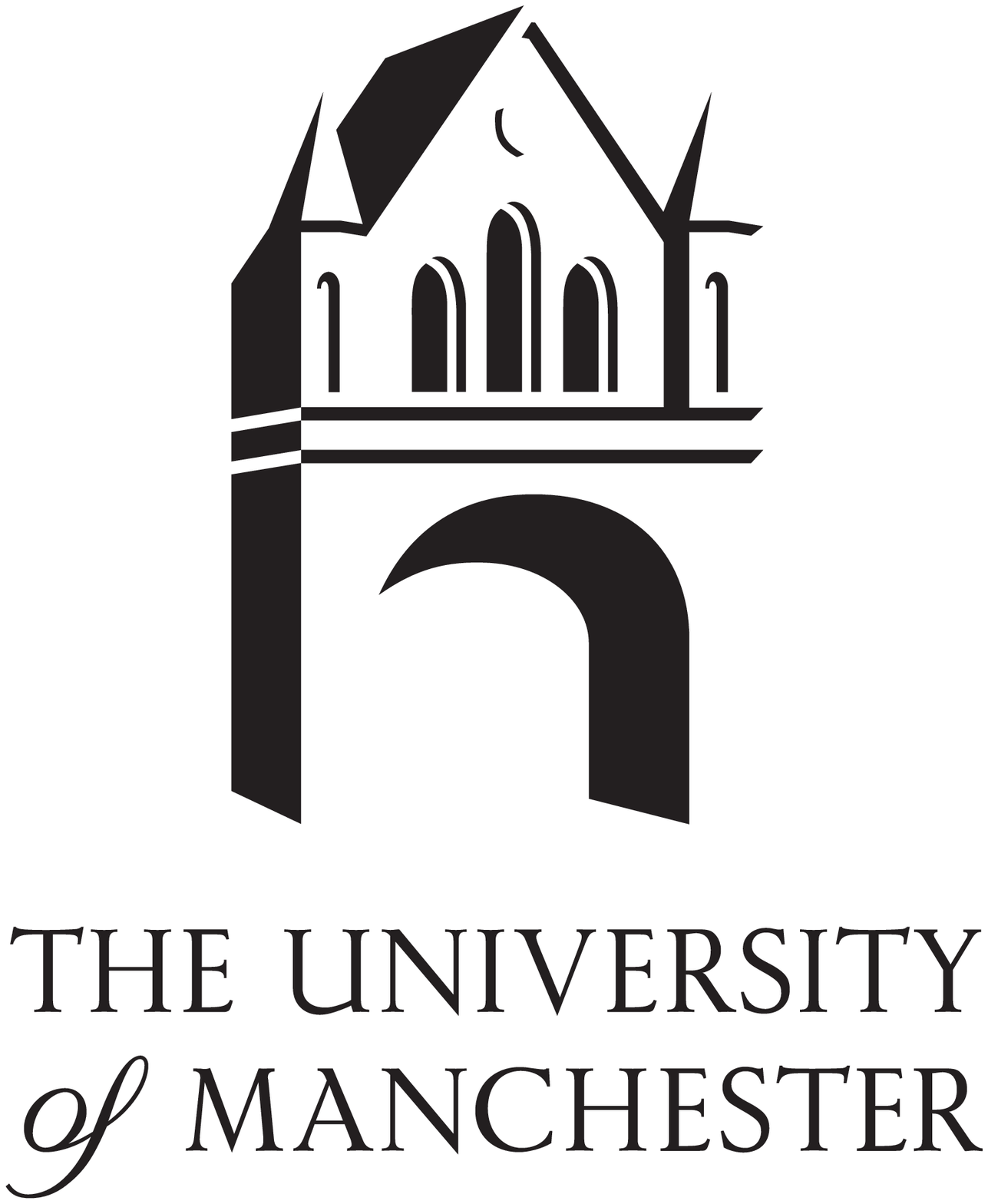} 
\hfil} \vfil }}

\def\header #1\par{ \centerline{\it #1}\par}
\def \letterhead {
\voffset -10 mm
{\advance \hsize by 2 cm
\font\address=cmss12
\vskip -9.5 pt
{\address
\vbox {
\vskip 2 mm
\hbox to 6 cm {\hskip -1 cm Department of Physics and Astronomy\hfill }
\hbox to 6 cm {\hskip -1 cm The University of Manchester \hfill }
\hbox to 6 cm {\hskip -1 cm Manchester \hfill }
\hbox to 6 cm {\hskip -1 cm M13 9PL \hfill }
\hbox to 8 cm {\hskip -1 cm Tel 0161-275-\phonenumber\hskip 1.0 cm 
Fax 0161-273-5867\hskip 1.0 cm}
%\hbox to 3 cm {\special {psfile=[roger.tex]UK.ps hscale=20 vscale=20 
%voffset=-75}}
\hfill
\vbox{
\hbox to 6 cm{\includegraphics{ulogo.ps}} 
\vskip 0.5 mm
}
}
}}
}

\def \letter #1 {
\topskip 0 pt
\vsize 599 pt
\nopagenumbers
\letterhead
\vskip 1 mm

\vbox to 3.4 cm {\vfill #1 \vfill}
\vskip 5 mm

\hbox to 2 cm {\hskip -1 cm \leaders\hrule height .5 pt \hfill \hskip 1 cm}

\rightline {\number\day \
\ifcase\month\or January\or February\or March\or April\or
May\or June\or July\or August\or September\or October\or
November\or December\fi \ \number\year}

\vskip 5 mm
}

%Some common Addresses
\def \AddressRAL #1 { \leftline{#1 }
\leftline{HEP Division,}\leftline{Rutherford Appleton Laboratory,}
\leftline{Chilton,}\leftline{Didcot,}\leftline{Oxon}}

\def \AddressRegistrar  #1 {
\leftline{#1 }
\leftline{The Registrar's Department}
\leftline{Main Building}
\leftline{University of Manchester}
\leftline{Oxford Road}
\leftline{Manchester M13 9PL}}

\def \AddressFaculty #1 {
\leftline{#1 }
\leftline{The Faculty of Science}
\leftline{Roscoe Building}
\leftline{University of Manchester}
\leftline{Oxford Road}
\leftline{Manchester M13 9PL}}

\def \AddressPhysics #1 {
\leftline{#1 }
\leftline{Department of Physics}
\leftline{University of Manchester}
\leftline{Oxford Road}
\leftline{Manchester M13 9PL}}

\def \AddressCERN #1/#2 {
\leftline {#1}
\leftline {#2 Division,}
\leftline {CERN,}
\leftline {CH1211 Gen\`eve 23,}
\leftline {Switzerland}}

\def \NIM #1 {{\it Nucl. Instr \& Meth. \/}{\bf A#1}\ }
\def \ZPC #1 {{\it Zeit. Phys. \/}{\bf C#1}\ }
\def \NPB #1 {{\it Nucl. Phys. \/}{\bf B#1}\ }
\def \PLB #1 {{\it Phys. Lett. \/}{\bf B#1}\ }
\def \PL #1 {{\it Phys. Lett. \/}{\bf #1}\ }
\def \PRD #1 {{\it Phys. Rev. \/}{\bf D#1}\ }
\def \PRL #1 {{\it Phys. Rev. Lett.\/}{\bf #1}\ }
\def \PR #1 {{\it Phys. Rev. \/}{\bf #1}\ }
\def \CPC #1 {{\it Comp. Phys. Comm. \/}{\bf #1}\ }

\newcount \eqnumber
\newcount \fignumber
\newcount \highref

\def \eq #1{\global\advance \eqnumber by 1
\let \rrr=\eqnumber
\xdef #1{\the\rrr}
\eqno (\the\eqnumber)
}
\def \fig #1{Figure \global\advance \fignumber by 1 \the\fignumber
\let \rrr=\fignumber
\xdef #1{Figure \the\sss}
}

\newcount\refcheck
\newcount\thisrf
\def\references #1 {
\thisrf=0
\ifnum\refcheck=0
\else
\leftline{\bf References}
\fi
\input #1
\refcheck=1
}

\def\refiopstyle #1:#2;#3\par{
\relax
\ifnum\refcheck=0
\edef \rrr{#2}
\let #1=\rrr
\else
#2 
\ 
#3
\par
\fi
\relax
}

\def\refseq#1{\ifnum#1>\the\highref\global\advance\highref 1  
\ifnum#1>\the\highref
\message{Reference out of Sequence: expecting \the\highref got #1}
\highref=#1\fi\fi}

\def\reference #1:#2\par{\advance \thisrf by 1
\relax
\ifnum\refcheck=0
\let \sss=\thisrf
\edef \rrr{\the\thisrf\noexpand\refseq{\the\thisrf}}
\let #1=\rrr
\else
\the\thisrf
:\ 
#2
\par
\fi
\relax
}

\font\small=cmr8

\def \tickbox {\lower 2 mm \hbox{
\vbox  {\vskip .5 mm\hrule \hbox to 4 mm
{\vrule \strut  \hfill  \vrule}\vfill\hrule} }}

\newcount\secnum
\newcount\examplenum
\newcount\subnum
\newcount\subsubnum
\newcount\chapnum
\font \splash=cmssi17
%\font \author=cmssi17
\font \titlefont=cmss17 scaled \magstep2
%\font \bf=cmcsc10

\font \address=cmss12
\font \cf=cmbxsl10
\examplenum=0
\secnum=0
\subnum=0

\def\chapter #1 {
\advance\chapnum by 1 \secnum=0 \subnum=0 \subsubnum=0
\vfill
\hbox{\bf \quad Chapter \number \chapnum : #1}}
\def\lecture #1 #2{
\secnum=0 \subnum=0
\hbox{\centerline{\splash SLUO Lecture #1: #2}}
\headline={\ifnum\pageno>1 SLUO Lecture #1 \dotfill #2 \fi}
\footline={\ifnum\pageno>1 \hss --\  \folio \ -- \hss  \else  \fi}
}

\def\subsection#1 \par{\par \advance\subnum by 1
\subsubnum=0
\goodbreak \vskip 0.3cm\leftline{\cf \number \secnum .\number 
\subnum \ #1} \par}

\def\subsubsection#1 \par{\par \advance\subsubnum by 1
\goodbreak \vskip 0.3cm\leftline{\sl \number \secnum .\number 
\subnum .\number \subsubnum \ #1} \par}

\def\section#1\par{\goodbreak \par \advance\secnum by 1 \subnum=0
\vskip 0.5cm\leftline{\bf \number \secnum . \ #1}\vskip 0.02cm \par
\message{ Section  \number \secnum    #1}
}

\long\def\example #1 {\par \advance\examplenum by 1
\vskip 12 pt \goodbreak \boxit {{\bf Example \number\examplenum :} #1 }
}

\long\def\boxit #1{\vbox {\kern-5pt\hrule\hbox{\vrule\kern3pt
           \vbox{\kern3pt #1 \kern3pt} \kern3pt \vrule} \hrule}}

%\settabs 5 \columns
\def\bull#1\par {\item {$\bullet$} #1 \par}

%%
% END OF STANDARD MACRO STUFF
%%

\docnum MAN/HEP/2002/1
\def \em {\it}
\def \bull {\item{$\bullet$}}

\def \subtitle #1\par{\vskip 6 pt \leftline{\hskip \leftskip \big #1}\par}
\def \example{{\bf Example: }}
\def \myth #1\par{{\bf Myth: #1}\par}
\font \csc=cmcsc12
\def \babar {{\csc BaBar}}

\title{Systematic Errors: Facts and Fictions}
\presentedat { Advanced Statistical Techniques 
in 
HEP, Durham, March 2002}
\author{Roger Barlow}

%\institute{Department of Physics and Astronomy, Manchester University, England}
\abstract{
The treatment of systematic errors is often mishandled.
This is due to lack of understanding and
education, based on a fundamental ambiguity as to what is
meant by the term. This note addresses the problems and
offers guidance to good practice.
}

\maketitle
\section{RIVAL DEFINITIONS: UNCERTAINTY AND MISTAKE}

\subsection { Random Uncertainties and Mistakes}

The word {\em error} is used in several ways.
The everyday dictionary definition is a 
synonym for {\em mistake}. 
In statistics this usage continues (as in `Type I error' for 
the 
rejection of a true hypothesis and `Type II error' for the acceptance of a false one) 
but it is also used in the sense of {\em discrepancy}: the
statistician writes the equation of
a straight line fit as $y_i= m x_i + c + \epsilon_i$
where $\epsilon_i$ 
is the `error term', the difference between the measured and the ideal value.

A physicist does not use this language; their interest is concentrated not on 
the actual discrepancy of a single measurement, but on the overall 
{\em uncertainty}. 
They write a straight line as $y_i= m x_i + c $
where the equals sign signifies agreement to some uncertainty (or resolution)
 $\sigma$,
and they will call this the `error'.
(In early texts this was called the `probable error' but the 
`probable' got dropped.)  This use of `error' to mean
`uncertainty' rather than `mistake' or `discrepancy' is common in the
language, for example
`error bar', `error analysis', and `quoted error
on the result.'

Suppose a set of measurements have been made of the same quantity, and the
values are

$$1.23, 1.25, 1.24, 1.25, 1.21, 1.52, 1.22, 1.27$$

These exhibit some {\it uncertainty} in the 3rd decimal place,  
and a {\it mistake} in that one of the values clearly does not belong 
with the others.

Statistics provides tools to identify and use the uncertainty. It can be 
estimated from the rms deviation of the values about the mean,
and then used in specifying the accuracy of this mean, or of a single
measurement, or the number of measurements that would be needed to
achieve some desired accuracy, and so on.

Statistics provides tools to identify a mistake, but not to use it.
We can see that the value of 1.52 is wrong - or, more correctly, that
the probability of this value being produced by a 
measurement consistent with the others
is so small that we reject it. Statistics does not
and cannot tell us what to do next. Perhaps the value is a transcription
error for 1.25. Perhaps it was taken while the apparatus was still warming up.
Perhaps it is due to an unforeseen and Nobel-prize-winning effect.  
Perhaps it is right and the others are wrong.
What we do next has to be based on experience and common sense, but
statistics does not prescribe it. 

\subsection { Systematic  Uncertainties and Mistakes}

For consistency
 physicists must use {\it systematic error} in the same way
as {\it random error}: 
to denote a {\it systematic uncertainty} and not a {\it systematic mistake}.
But consider the following two definitions

`{\it Systematic effects} is a general category which includes effects 
such as background,
selection bias, scanning efficiency, energy resolution, angle resolution,
variation of counter efficiency with beam position and energy, dead time, etc.  The uncertainty in the estimation of such a systematic effect is called a 
{\it systematic error}.'
\hfill - Orear[1]
 
`{\it Systematic Error:\/} reproducible inaccuracy introduced by faulty equipment,
calibration or technique.' \hfill - Bevington[2]\break

These are taken from widely read and accepted authors, and each on its own 
would probably get a nod of approval from a practising physicist.  However
putting them together shows that they are incompatible.  The
second definition means mistake - the word `faulty' is a key.
The first explicitly  defines an uncertainty.  It does not contain the
sense of fault, blame, or incompetence which is fundamental to the second.

The following examples of `systematic error' show these two usages.

\item 1
The energy $E$ measured in a calorimeter module is given by
$$E= \alpha D + \beta,$$
where $D$ is some digitisation of the recorded output signal.  The 
error (=uncertainty) on $E$ has a random part due to the random uncertainty
on $D$ (from sampling statistics). It has a systematic part due to
errors (=uncertainties) on the calibration constants 
$\alpha$ and $\beta$. These are systematic in that from measurement to measurement the
value of $D$ will fluctuate about its true value with some
standard deviation $\sigma_D$, whereas the values of $\alpha$ and $\beta$
are constant and their discrepancy is applied systematically to all
measurements.
\item 2 
A branching ratio $B$  is calculated from number of observed decays 
$N$ out of some total number $N_T$, where the efficiency
of detection is $\eta$: $$B= N/(\eta N_T).$$
There is a statistical Poisson (or perhaps binomial) error on the ratio
$N/N_T$ which will fall as more data is gathered. 
There is an uncertainty on the efficiency $\eta$ (probably calculated from Monte Carlo simulation) whose contribution will not
(unless other steps are taken) fall as more data is taken.

\item  3 
Measurements are taken with a steel rule.  The rule was calibrated
at a temperature of $15$ C and the measurements are taken in a warmer laboratory, and
the experimenter does not allow for thermal expansion.

\item 4  During the processing of data, numbers are rounded down by omitting
all digits after the decimal point.

The first two examples are systematic errors in Orear's sense. There is a 
systematic effect, encapsulated in $\alpha$, $\beta$, and $\eta$, and an
uncertainty in that effect, encapsulated in 
$\sigma_\alpha$, $\sigma_\beta$, and $\sigma_\eta$.
These errors can be handled by standard techniques, as will be 
described later.

The third and fourth are examples of the second definition,
indeed Example 3 is taken from Bevington; they arise
from mistakes.  In order to consider how to handle them one has to 
specify the situation more precisely (as will be done in what follows.)

For consistency we should use
Orear's definition rather than Bevington's.
In an ideal world the term `systematic error' might be avoided,
and replaced by `systematic uncertainty' but that is unrealistic. 
It is vital to distinguish {\it systematic effects}
from the {\it systematic errors} which are the uncertainties 
in those effects and
from the {\it systematic mistakes} resulting from the neglect of such
effects. Confusion between these three concepts is widespread, and
responsible for poor practice. 

Of course 
systematic mistakes still exist, and
still need to be identified.
But calling them {\em mistakes}  makes clear that although statistics 
can help to find them, it
does not provide tools to tell us what to do with them.

\subsection{ Systematic Errors and Bias}

The terms `bias' and `systematic error' are 
treated as synonymous by some authors [3,4,5].
This is not a full enough definition to be helpful.  In discussing a
bias one has to consider its status.  

Once a bias is known,
it can be corrected for: an estimator with known bias can be
trivially replaced by an unbiassed estimator.
If the bias is unknown and unsuspected then one can by
definition do nothing about it.
The match between `bias' and  `systematic error' under our definition is
the case where a bias is known to exist, but its
exact size (systematic effect) is unknown (systematic uncertainty).

We apply this to the example of measurements with an expanding steel rule.
\item 1 
If the expansion coefficient is known, as are the two temperatures of
calibration and actual measurement, then the measurements can be
corrected and the bias is removed; the systematic effect is known exactly
and there is no systematic error.  
\item 2 If the effect is ignored then this is a
mistake. Hopefully consistency checks will be done and will (through
statistical techniques) reveal
a discrepancy for which the physicist will (through common sense,
experience and intuition) realise the cause.
\item  3 If the effect is known to exist but 
the temperature at which the actual measurements
was taken was not recorded, and one can only give the 
laboratory temperature to within a few degrees, that is a 
systematic uncertainty on a systematic effect, and a systematic error in
the accepted sense.

\section{ SYSTEMATIC ERRORS CAN BE BAYESIAN}

A {\it random} uncertainty fits neatly into the frequentist definition
of probability.  In considering a large ensemble of measurements,
different results are obtained. One can speak of the probability
of a particular result as the limit of a fraction  
of measurements giving that result.
But
if a measurement with a {\it systematic} uncertainty is
repeated
then, by definition,
the same result is obtained every time, 
giving an ensemble of identical results which cannot be used to say
anything about probability.

In some cases there is a clear way out.  The calibration of a
calorimeter, for example, may be determined through a
separate experiment; the ensemble to be considered is then
the ensemble of calibration experiments, rather than the ensemble of
actual measurements.  A resistor with value $100 \pm 10\ \Omega$ used in 
voltage and current measurements will always have the same value
(of, perhaps, $106\ \Omega$) but it came from a drawer full of
nominal $100\ \Omega$ resistors with a spread of values.

In some cases there is no escape. This occurs particularly for
so-called `theory errors'.   For example, consider the determination
of luminosity in  $e^+e^-$ collisions
 through measuring small angle Bhabha scatters.
Perhaps the cross section has been calculated 
to third order in the fine structure constant $\alpha$.
It is inaccurate in that it deviates from the exact expression.
Yet a different calculation will always gives same
result.
One can guess at this inaccuracy: setting it to a few times
$\alpha^4$ would be sensible. But there is no
(obvious) ensemble to use. To quote a figure for an uncertainty
in such a situation requires one to use a subjective (Bayesian) definition
of probability.

Even for a practitioner who generally uses and advocates a frequentist
definition of probability, there are times when the Bayesian definition
has to be invoked. This can be excused when a particular systematic error
is (as it usually is) a small part of the total error.
In doing so it is important that one is aware of
what one is doing, and the possible pitfalls. 

\subsection {Prior pitfalls: an illustration}

These dangers appear in a recent example [6].
Consider an experiment where 
limits are obtained on some quantity $R$ (perhaps
a branching ratio or cross section) from some observed number of events $n$.
This was considered by
Cousins and Highland [7] who wrote 
 $$n = S R$$
where $S$ is the `sensitivity' factor, containing factors such as the
detection efficiency, and therefore has some associated uncertainty
$\sigma_S$ which is probably Bayesian.  
The limits on $R$ are compounded from the statistical
(Poisson) variation on $n$ and the variation in $S$.  Consider a particular
value of $R$ as an upper limit:
the confidence level can be computed by repeatedly taking that
value, multiplying it by a value drawn from $Gauss(S, \sigma_S)$,
and using that as the mean for generation of a Poisson random number.
The fraction of times that this value is less than or
equal to the observed $n$ gives the confidence level for this
value of $R$. $R$ can then be adjusted and the process repeated
till the desired confidence level is attained.
This can be done using approximate analytical formulae [7] or by a
toy Monte Carlo [6,7]

 However it would be equally valid to write [8]
$$
R = A n $$
where the appropriate factor $A$ is merely the inverse of $S$.
A trivial change.  And yet if one applies the same proportional
uncertainty to $S$ and to $A$ one gets different results.
For example, suppose 3 events  are observed, 
and you have an
uncertainty of 10\% on $S$ or $A$, which are both taken as 1,
and consider $R=5$.
The probability of 3 events or less is 
27.2\% from the first method but 26.6\%
from the second. 
The results are different because
the priors are different; a Gaussian in $S$ is not the same as a
Gaussian in $A\equiv 1/S$.

A third possibility would be to use a
Jeffreys' prior. The prescription for this
is to effectively transform to a variable for which the
Fisher information is flat, and take a flat prior in that. 
To call this `objective' is an
overstatement, but it does offer a unique
prescription. Here it means working in $\ln A$  
or equivalently $\ln S$, and generating a Gaussian in that.
This gives a value intermediate between the two others.

The moral is that, as is well known to statisticians, with
Bayesian statistics one must
(unless one has some {\em a priori} reason for a
particular form)  
investigate the stability of a result under changes in the prior.
This example shows that variation does occur at the sort of level
to which results are generally quoted.

\vfill\eject

\section {EVALUATING EFFECTS OF SYSTEMATIC UNCERTAINTY}

\vbox to 6 cm
{
\includegraphics{eplot.ps}
}

\centerline{Figure 1: Evaluating the effect of an uncertainty}

There is a widespread myth that when errors are systematic,
the standard techniques for errors do not apply, and practitioners
follow prescriptions handed down from supervisor to student.
This is not so.
The standard undergraduate 
combination of errors formula still applies, though
one has to be careful to include the correlation terms.

In some cases this is all that is required. If the energy measurement
has a systematic error then the error on, say, an invariant mass
made from these quantities can be found in the standard way.
In other cases they cannot.
Suppose the experimental result $R$ depends on some parameter
$a$ which is not known exactly, but with some uncertainty
$\sigma_a$.  This parameter $a$ could be one that
affects the Monte Carlo generation of 
simulated events used to extract other quantities in the analysis, 
which means that
the effects of this uncertainty cannot be followed through
combination of errors algebra. Instead one generates samples
at the best value  $a_0$, and usually at two other values, 
$a_0+\sigma_a$ and $a_0-\sigma_a$ to obtain $R'={dR \over da}$,
as shown 
in Figure 1.
The quoted result is $R(a_0)$, and the error due to the
uncertainty in $a$ is $\sigma_a R'$ which is the difference in $R$. 
(In some cases more points may be appropriate to investigate possible
non-linearity, or  different $a$ values to avoid numerical errors. The
choice to evaluate at $\pm \sigma$ is for convenience.)
This can be done for the final result or for
some intermediate result which can then be used in a combination of
errors formula. 
Indeed with today's processing power this method is generally used
rather than using algebra, as it gets straight to the answer without 
assumptions about effects being small.

In some cases this procedure can be
simplified: for example if the invariant mass
is used to select pairs of photons in a window near the $\pi^0$ mass,
and the number of these pairs 
used to give a further value,
then given an uncertain energy scale,  one can
vary the window
rather than the energy scale
 by the appropriate amount, 
 and redo only the final
part of the analysis.  Note (for future reference) that in such a case the upper and lower
edges of the window are varied together, coherently, and that they
are changed by a prescribed amount.

\subsection{ Evaluation: the error on error paradox}

\vbox to 6 cm
{
\vfill
\includegraphics{eeplot.ps}

\centerline{Figure 2 Errors on errors}

}

In a typical experiment there may be a large amount of Monte Carlo
data generated 
at the central value $a_0$, but less at $a_0 \pm \sigma_a$.
So the estimate of $R'$ may itself have an error, $\sigma_{R'}$,
due to
finite Monte Carlo statistics.
How does this affect the systematic uncertainty on $R$?
There are 3 suggestions.

\item 1  $\sigma^2=(R' \sigma_a)^2 + (\sigma_{R'} \sigma_a)^2$

The uncertainty in $R'$ is another uncertainty 
so it should be added in
quadrature.

\item 2  $\sigma^2=(R' \sigma_a)^2 - (\sigma_{R'} \sigma_a)^2$

This value $R'$ has been modified from true $R'$ in such a way that
 $\langle R'^2 \rangle $ is increased. 
(If  R is independent of $a$,  these
errors will force $R'$ away from zero.) Subtraction in 
quadrature compensates for this.

\item 3  $\sigma^2=(R' \sigma_a)^2$

There is no point messing about with such subtleties. This 
correction is going to be small unless both $\sigma_a R'$ and
$\sigma_R'/R'$ are large. 
In that case then to do a decent job on the measurement you
have to go back and generate more Monte Carlo data.

\subsubsection{An illustrative example}

If the estimate of $R'$ varies symmetrically about its true value, and
there is no reason to doubt that it does, then
the estimate of $\sigma$ as $R' \sigma_a$ is unbiassed.
The problem is that this estimate is
incorporated in the total error by addition in quadrature; what matters is
not the standard deviation but the variance. And if our
estimate of $\sigma$ is unbiassed then our estimate of $\sigma^2$ is
certainly not unbiassed.

To avoid some of the unnecessary complication we consider an
illustrative  example. Suppose an integer $x$ is generated with uniform 
probability over a large range.
It is then either increased or decreased by 1, with equal
probability, to give $y=x \pm 1$.
You are given the value $y$, and
you need the best estimate of $x^2$.
(This represents the need to know the variance rather
than the standard deviation.)

There is a (Bayesian) viewpoint which argues: suppose $y$ has
a particular value, say $y=5$. This could have come from $x=6$ or $x=4$,
and the probabilities are (by symmetry, ignorance, etc) equal.
Your value of $y^2$ is 25, but the true value is 16 or 36. The midway
point is 26, and that value will be unbiassed. So add 1 to the
value of $y^2$.
This is the first of the 3 methods above.

There is a (frequentist) viewpoint which argues in the reverse
direction.  Suppose $x$ has a particular value, say $x=5$. This
could give $y=4$ or $y=6$ with 
equal probability.
The true value of 25 becomes 16 or 36.
On average this is 26, so subtract 1 to remove the bias.
This is the second of the 3 methods above.

Algebraically, these two arguments take the two equations
$$y=x \pm 1 \qquad x = y \mp 1$$
square them to get
$$y^2=x^2 \pm 2 x + 1 \qquad x^2 = y^2 \mp 2 y + 1$$
and then argue that one can take the average, which cancels 
the $\pm$ or $\mp$ term.

We can test these arguments against the special case of zero.
Suppose you are given a $y$ of 0.
Argument 1 gives $0^2+1=1$.   Which is spot on, as we know
$x=-1$ or $+1$ so $x^2=1$ either way.
Argument 2 gives $0^2-1 = -1$, which looks crazy. 
How can our `best' estimate of $x^2$ be a negative number? 

Continuing the testing, suppose you generate $x=0$.  This will give
$y^2=1$ so argument 2 is spot on and argument 1 is out by 2.
Argument 1 will never give 0.
So argument 2 wins this test, but not so dramatically, as
you can never know whether $x$ was zero, but if $y$ is zero
this is obvious. 

In resolving a paradox one has to scrutinise the way it is posed, and
here the assertion of a `uniform probability over a large range' is
open to question; the nature of this prior in $x$  
affects the Bayesian argument 1 but not the frequentist
argument 2. There is no scale in the original problem
concerning $R'$, so a uniform probability up to a known finite limit
is inadmissable (and would introduce corrections at the limits).
You have some belief about the limits $\pm L$, and you believe $x$
is uniformly generated within these limits. This combines to
give a prior which falls off
at large $|x|$.
Your subjective probability of a result between
2 and 5 is larger than that for 10002 and 10005.
Given this fall, higher $|x|$  values are intrinsically  less probable than
low ones, so $y=5$ must be slightly more likely to have come
from $x=4$ than $x=6$. Any given $y^2$ value is more likely to be an
upward fluctuation than a downward one. 
This argument appears inescapable, in that it cannot be deemed to be small and
thus ignored. (If the fall in probability is very slow, then large values
are very probable and the size of the correction increases.)

Thus the logic of argument 1 fails, and we are left with argument 2.
This {\it is} the frequentist solution, and this {\it is} a valid 
frequentist problem: even if $\sigma_a$ has a Bayesian nature
the problem can be stated in terms of an ensenble in which the Monte
Carlo is rerun many times.
So it is technically correct. It gives the unbiassed estimate,
in the sense that averaged over many such estimates the bias will be 
zero.
\subsubsection {Conclusions for  errors on errors}

There is thus no justification for adding in quadrature, and there is 
a possible argument for subtraction. 
 But to do this requires  that measurements
with $\sigma_{R'}>|R'|$ must contribute negatively to the systematic
estimate, on the grounds this compensates for overestimation in 
other cases.
(And the greater the inaccuracy, the greater the reduction in the error.)
 To be right in general you may
have to do something manifestly wrong in an individual case (a feature well known
in frequentist confidence levels near boundaries.)   

If you have a large number of such corrections for
parameters $a,b,c...z$ then this approach may be arguable.
But not if
it's unique. You will never get it past the referee:
you investigate an uncertainty and as a result you {\em reduce} the
systematic error, on the grounds that you might have increased it
(or, perhaps, that in many parallel universes you increased it?)

No, at this point statistical sophistication has clearly gone too far
for plain common sense.   We therefore recommend Argument 3: that this
error on error correction should not be done as there is no
sensible way of doing it.  It can be left out of the reckoning if small, and 
if large
then more work is needed to make it small.

\section{CHECKS: FINDING MISTAKES}

Finding mistakes is done by thinking through the
analysis in a critical way, something often best done by
consulting colleagues or presenting results at seminars. Such a 
critique looks at what could go wrong, and at what checks can
be done on the analysis which could reveal mistakes.
These checks are variations of the analysis, for which
the correct outcome is known, either absolutely or in relation to
other results.

  You can
never prove that an analysis is perfect, but the more 
checks you perform successfully, the greater the credibility
of the result.

Such checks commonly include:

\item 1 Analysing separate data subsets

\item 2 Changing cuts

\item 3 Changing histogram bin sizes

\item  4 Changing parametrisations (including the order of polynomial)

\item 5 Changing fit technique

\item 6 Looking for impossibilities

This approach is shown, for example, in the \babar\ CP violation
measurement[9]

`... consistency checks, including separation of the data by decay mode, tagging category and $B_{tag}$ flavour... We also fit the samples of non-CP decay modes for $\sin 2\beta$ with no statistically significant asymmetry found.'

\subsection{ What is a significant difference?}

If an analysis is performed in two ways (say, using two
different forms to fit a background function)
then one hopes that the difference between the two
resulting values will be small; a large difference would suggest that
the background subtraction was not being done properly.
However it would be unrealistic to expect them to be identical.
The question arises as to what `small' means in this context.

It does not mean `small with respect to the statistical error'. 
The statistical error is probably dominated by the sampling
process.  But these two analyses are done on the same data (or
their datasets share a lot of elements), 
and so should
agree much better than that.

Suppose the standard analysis gives $a_1 \pm \sigma_1$. 
A different method done as a check gives $a_2 \pm \sigma_2$
We consider the difference $\Delta = a_1-a_2$.   
The error on this is 
$$\sigma_\Delta^2 = \sigma_1^2 + \sigma_2^2 - 2 \rho \sigma_1 \sigma_2$$

Suppose firstly that the estimate is a mean of some quantity $x$,
and that the check consists of selecting a subset of the data

\vbox to 3.2 cm{
\includegraphics{subsamples.ps}
\vfill

\centerline{Figure 3. Many checks can be performed by analysing a selected
subset of the total data.}

}

The two values are given by
$$a_1 = {1 \over N_T} \sum_T x_i  \qquad a_2={1 \over N_S} \sum_S x_i$$
and the errors by
$$\sigma_1 = {\sigma \over \sqrt {N_T}} \qquad  \sigma_2 = {\sigma \over \sqrt {N_S}}$$
and  the covariance between them is
$$  Cov(a_1,a_2) = N_S {1 \over N_T} \, {1 \over N_S} \sigma^2$$
so the correlation is just
$$ \rho = \sigma_1/\sigma_2 .$$
This gives the required error on $\Delta$
$$\sigma_\Delta^2 = \sigma_2^2 - \sigma_1^2 $$
showing that the error is found by subtraction in quadrature
of the two separate errors.

If the check is more general, perhaps using a different
method on the same data, it is still true that
$$\sigma_\Delta^2 = \sigma_1^2 + \sigma_2^2 - 2 \rho \sigma_1 \sigma_2.$$

The correlation $\rho$ is not known, but limits can be placed on it [10].
Introduce (briefly) a weighted average $$a(w)=w a_1 + (1-w) a_2.$$ 

This has variance
$$\sigma_{a(w)}^2 = w^2 \sigma_1^2 + (1-w)^2 \sigma_2^2 + 2 w (1-w) \rho \sigma_1 \sigma_2$$

By choosing $w$ (differentiate the above, set it to zero, solve for
$w$ and put back in) one gets the smallest variance possible from
a weighted sum. 
$$ \sigma_{min}^2 = {\sigma_1^2 \sigma_2^2 (1-\rho^2) \over \sigma_1^2+\sigma_2^2
- 2 \rho \sigma_1\sigma_2}.$$

Now, in an estimation problem there is a limit on the 
variance of any estimator: 
the Minimum Variance (or Cramer-Rao) Bound.
This limit applies
irrespective of the estimation technique used. It depends only on the
likelihood function, and its value $\sigma_0$ can be calculated from it.   
This bound means that
$$ \sigma_{min}^2 \geq \sigma_0^2.$$

Inserting the expression for $\sigma_{min}^2$ gives an expression
which can be rearranged to give limits on $\rho$, and that translates
to limits on
$\sigma_\Delta$.
$$\sqrt{
\left( \sigma_1^2 - \sigma_0^2\right)} +
\sqrt{\left( \sigma_2^2 - \sigma_0^2\right)}  \geq \sigma_\Delta \geq
\left|\sqrt{\left( \sigma_1^2 - \sigma_0^2\right)} -
\sqrt{\left( \sigma_2^2 - \sigma_0^2\right)}  \right|.$$

Notice that if $\sigma_1 = \sigma_0$ this again gives
subtraction in quadrature.  In many cases the standard analysis will be
the most efficient possible, so this will be the case.

\subsection {Checks that pass, checks that fail}

The standard procedure for doing an analysis can be caricatured
as follows

\item 1
Devise cuts, get result.

\item 2
Do analysis for random errors (likelihood or
   Poisson statistics.)

\item 3
Make big table.

\item 4
Alter cuts by arbitrary amounts, put in table.

\item 5
Repeat step 4 until time/money/supervisor's patience
is exhausted.

\item 6
Add variations in quadrature.

\item 7
Quote result as `systematic error'.
 
\item 8
If challenged, describe it as `conservative'.

This combines evaluation of errors with checks for mistakes, in a totally
inappropriate way.

Suppose a check is done, and a discrepancy emerges as some number 
of $\sigma_\Delta$. You then have to decide whether it has passed or failed
the test.  Your decision will depend on the size of the discrepancy
(less than 1 $\sigma$ surely passes, more than 4 $\sigma$ surely
fails), the number of checks being done (if you do 20 checks you expect
on average one $2 \sigma$ deviation) 
and at some level on the basic plausibility and reasons that
motivated the check (you might accept that data
taken in the summer were different more readily than you would accept that
data taken on Tuesdays were different.)

If  a check  passes then the correct thing to do is
{\em nothing}.  Put a tick in the box and move on.
Do not, as is practice in some areas, add the small discrepancy to 
the systematic error.  

\item 1 It's an inconsistent action. You asked `is there an effect' and
decided there wasn't. If there was no effect then you should not allow for it.
Remember that this is a check and not an evaluation of an effect.

\item 2 It penalises diligence.  The harder you work and more thorough you
are, the bigger your systematic error gets. A less careful analysis will
have a smaller quoted error and get the citations.

\item 3  Errors get inflated.   Remember how the LEP experiments 
appear to agree with each other and the 
Standard Model far {\it too} well.

One has to be careful. Contrast moving mass cuts by a defined amount
to compensate for energy uncertainty
(this is an 
evaluation and included) and changing mass cuts by an arbitrary amount
to check efficiency/purity
(this is a check and not included if successful.)

If it fails then the correct actions  to take are

\item 1
Check the test. The mistake may well lie there. Find and fix it.

\item 2 If that doesn't work, check the analysis. Find and fix mistake.

\item 3 Worry. Maybe with hindsight an effect is reasonable. (Why are the
results of my ruler measurements different after lunch? Hey, the temperature's
warmer in the afternoons - I forgot about thermal expansion!)
 This check now becomes
 an evaluation.

\item 4 Worry. This discrepancy is only the tip of the iceberg. Ask
colleagues, look at what other experiments did.

\item 99 As a last resort, incorporate the discrepancy
in systematic error.

Just doing a whole lot of checks and
adding up the results in quadrature to the systematic error is making a
whole lot of mistakes, some too lenient, some too harsh.  

\subsection {Illustration: an inappropriate function}

\vbox to 5 cm
{
\vfill
\includegraphics{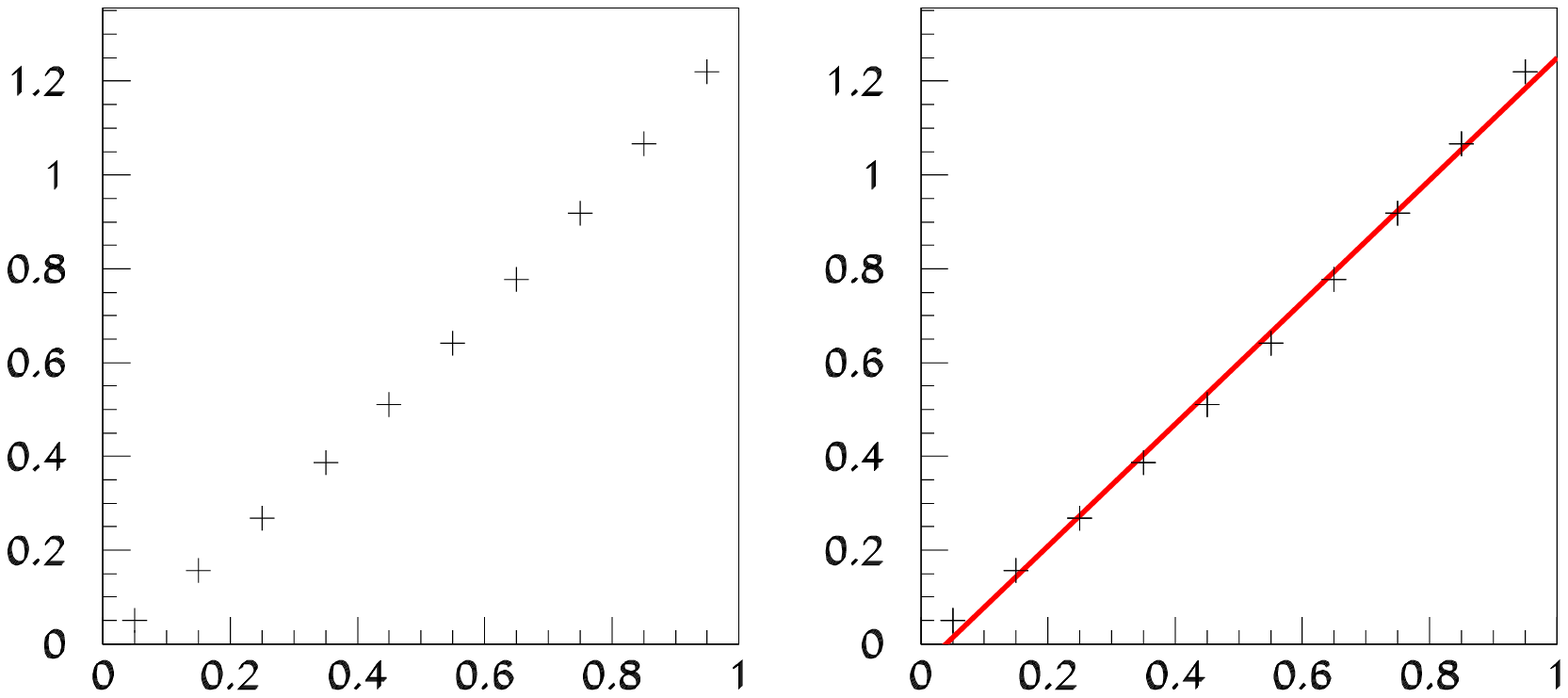}
\includegraphics{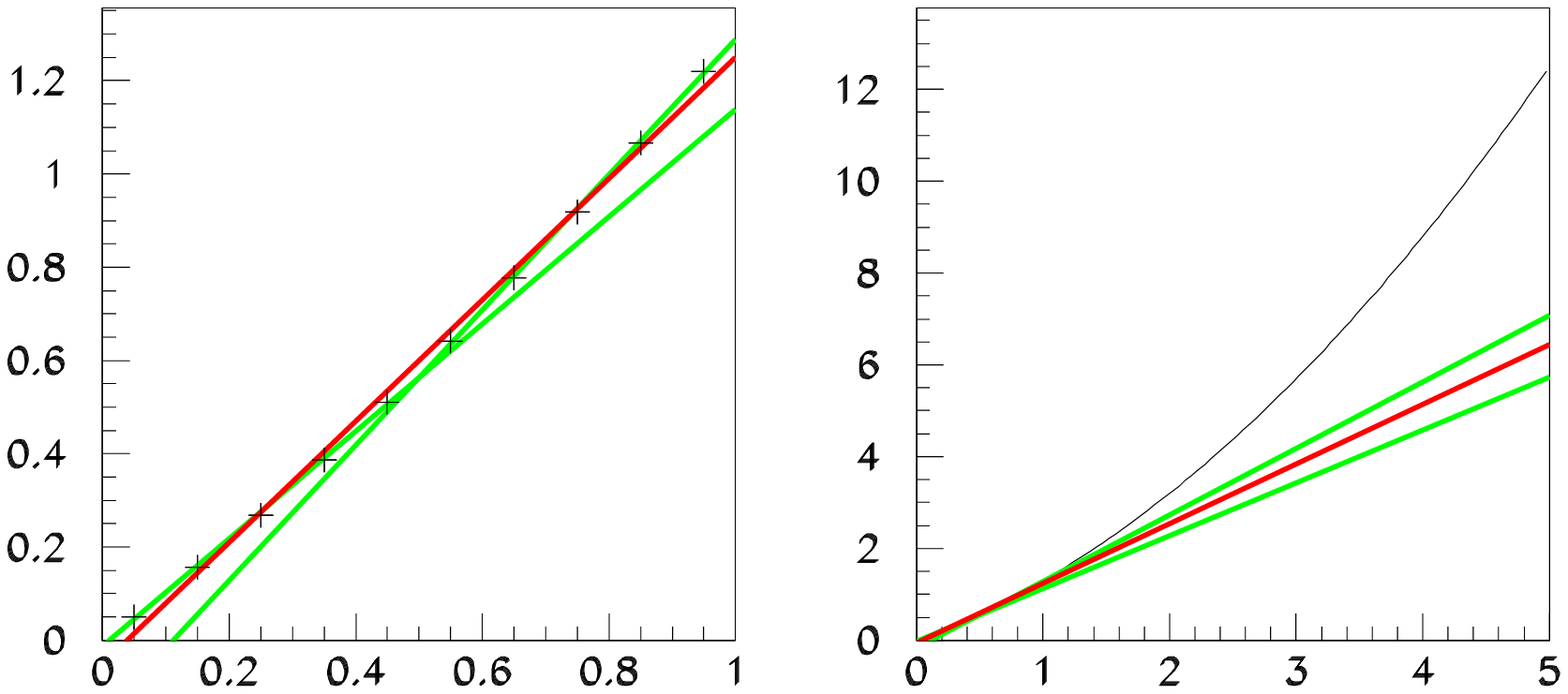}
\centerline{
Figure 4 An inappropriate function. The plots show (i) The data. 
}\centerline{
(ii) A
straight line fit to the data. 
(iii) Additional fits to subranges. }
\centerline{(iv)
Extrapolation of fits and the true function - note changes in scale.
}
}
 
Suppose you are using a calorimeter for which the energy
$y$ corresponding to a signal $x$ is actually given  by $y=x+0.3 x^2$. 
Measurements are taken as shown in the first plot (measurement errors are
suppressed for clarity).
You fit it as a straight line $y=mx+c $
using data in the range $0\leq x \leq 1$,
and get $m=1.3$ and $c=-0.05$, as shown in the second plot. 
This is what you 
use in your analysis.

As a sensible check you decide to calibrate the subranges
$0\leq x \leq 0.5$ and $0.5 \leq x \leq 1$ separately.
The results are different (as shown in the third plot). The slopes
are 1.15 and 1.45, and there is no possibility that this is 
a statistical error.

You follow the procedure above but for some reason
fail to spot that a linear calibration is inadequate. 
You end up incorporating the difference of 0.15 as a systematic error on $m$
(with perhaps a similar systematic error for $c$, and even a 
correlation between them.)

Notice what a terrible decision this is. As you can see from the
figures, in the range $0\leq x \leq 1$ this is far too harsh.
The line with a slope of $1.3$ actually follows the points
pretty well and this extra error is inflationary.

On the other hand, if this calibration is to be extrapolated 
towards $x=2$ or even $x=5$, then even this extra variation far
underestimates the calibration discrepancy in this region. The
procedure is far too lenient. 

This illustrates the point that there 
is no `correct' procedure for incorporation of a check that fails.
If you fold it into the systematic errors this is almost certainly
wrong, and should only be done when all other possibilites have
been exhausted.

\section { CONCLUSIONS: ADVICE FOR PRACTITIONERS}

The following should be printed in large letters and hung on the wall of every
practising particle physicist.

\item I Thou shalt never say `systematic error' when thou meanest
`systematic effect' or `systematic mistake'.%lest thou be justly smitten by thunderbolts.

\item {II} Thou shalt not add uncertainties on uncertainties in quadrature.
If they are larger than chickenfeed thou shalt generate more
Monte Carlo until they shrink to become so.
 
\item {III} Thou shalt know at all times 
whether what thou performest is a check for a mistake
or an evaluation  of an uncertainty.

\item {IV} Thou shalt not incorporate successful check results
into thy total systematic error and make thereby a shield behind which
to hide thy dodgy result.

\item V Thou shalt not incorporate failed check results unless thou art
truly at thy wits' end.

\item {VI} Thou shalt say what thou doest, and thou shalt be able
to justify it out of thine own mouth; not the mouth of
thy supervisor, nor thy colleague who did the analysis last time,
nor thy local statistics guru, 
nor thy mate down the pub.

\rightline{Do these, and thou shalt flourish, and thine analysis likewise.}
\vskip1cm
\section{ACKNOWLEDGEMENTS}

I would like to thank Louis Lyons, James Stirling, Mike Whalley and others
who organised a lively, different and much needed conference on this topic.
Also all those too numerous to mention with whom I have had 
enlightening discussions on this subject over the years.

\noindent

\section {References}

[1] J. Orear, Notes on Statistics for Physicists, UCRL-8417,\hfill\break {\tt http://nedwww.ipac.caltech.edu/level5/Sept01/Orear/frames.html}

[2] R. Bevington, Data reduction and Analysis for the Physical
Sciences. McGraw Hill 1969

[3]  B. L. van der Waerden, Mathematical Statistics, p160, Springer 1969.

[4] M. G.  Kendall \& W. R. Buckland, A Dictionary of Statistical Terms,
3rd Edition, Oliver and Boyd, 1971.

[5] R. A. Fisher, Statistical Methods For Research Workers, p207,
14th Edition, Oliver and Boyd, 1970

[6] R. J. Barlow, A Calculator for Confidence Intervals,
hep/ex-02030002, to be published in Comp. Phys. Comm.

[7] R. D. Cousins and V. L. Highland, Nucl. Instr \& Meth. {\bf A320} p331, 1992

[8] The BaBar Statistics Working Group, 
Recommended Statistical Procedures for BaBar, BaBar Analysis Document
\# 318,
 (Oct18 2001)
\hfill\break 
{\tt http://www.slac.stanford.edu/BFROOT/www/Statistics/Report/report.ps}

[9] The BaBar collaboration (B. Aubert et al.),
Measurement of CP-Violating Asymmetries in $B^0$ decays to CP eigenstates,
 Phys. Rev. Lett. {\bf 86}, p2515, 2001

[10] M. G. Kendall and A. Stuart, The Advanced Theory of Statistics,
Section 17.27, Vol II, 4th 3 volume edition, Charles Griffin and Co, 1979

\bye